\documentclass[a4paper,12pt]{article}
\usepackage[left=2.2cm,right=2.2cm,top=2cm,bottom=2.5cm]{geometry}

\renewcommand{\arraystretch}{1.2} 
\usepackage{caption}
\usepackage[utf8]{inputenc}
\usepackage{multirow}
\usepackage[pdftex]{graphicx}
\usepackage{amssymb}
\usepackage{amsmath}
\usepackage{cite}
\usepackage{braket}
\usepackage{slashed}
\usepackage{longtable}
\usepackage{booktabs}
\usepackage{array}
\usepackage[normalem]{ulem}
\usepackage{subcaption}

\usepackage[dvipsnames]{xcolor}
\usepackage{xspace}
\usepackage{makecell}

\usepackage[]{hyperref}
\allowdisplaybreaks
\numberwithin{equation}{section}

\newcommand{\refapp}[1]{\hyperref[app:#1]{Appendix~\ref*{app:#1}}}
\newcommand{\reffig}[1]{\hyperref[fig:#1]{Fig.~\ref*{fig:#1}}}
\newcommand{\refeq}[1]{\hyperref[eq:#1]{Eq.~(\ref*{eq:#1})}}
\newcommand{\refeqs}[2]{\hyperref[eq:#1]{Eqs.~(\ref*{eq:#1})-(\ref*{eq:#2})}}
\newcommand{\refeqa}[2]{\hyperref[eq:#1]{Eqs.~(\ref*{eq:#1})} and~\hyperref[eq:#2]{(\ref*{eq:#2})}}
\newcommand{\refsec}[1]{\hyperref[sec:#1]{Section~\ref*{sec:#1}}}
\newcommand{\refsubsec}[1]{\hyperref[sec:#1]{Subsection~\ref*{sec:#1}}}
\newcommand{\reftab}[1]{\hyperref[tab:#1]{Table~\ref*{tab:#1}}}

\renewcommand{\Im}{\text{Im}}
\newcommand{\MeV}{\text{MeV}}
\newcommand{\GeV}{\text{GeV}}
\newcommand{\MSbar}{$\overline{\text{MS}}$\xspace}

\newcounter{TODO}

\newcounter{TODOd}

\begin{document}

\vspace*{-2mm}
\begin{center}
\fontsize{15}{20}\selectfont
\bf
\boldmath
Correlator with tensor currents and two masses at two loops
\end{center}

\vspace{2mm}

\begin{center}
{Terry Generet$^{\,a}$, Nico Gubernari$^{\,b}$, Eetu Loisa$^{\,b}$}\\[5mm]
{\it\small
{$^{\, a}$} 
Cavendish Laboratory, University of Cambridge, JJ Thomson Avenue,\\ 
Cambridge CB3 0HE, United Kingdom
\\[2mm]
{$^{\, b}$} 
DAMTP, University of Cambridge, Wilberforce Road,\\
    Cambridge, CB3 0WA, United Kingdom
\\[2mm]
E-mail: {\textnormal{\texttt{tg513@cam.ac.uk}, \texttt{nicogubernari@gmail.com}, \texttt{eal47@cam.ac.uk}}}
}
\end{center}

\vspace{1mm}
\begin{abstract}\noindent
\vspace{-5mm}

\noindent
We calculate the vacuum-to-vacuum correlator of two flavour-non-diagonal tensor currents of massive quarks, retaining the full dependence on the two quark masses and the external momentum.
For the first time, we include perturbative corrections up to next-to-leading order.
Our fully analytical expressions are provided in machine-readable form.
Furthermore, we present numerical results for various input parameters, including an estimate of the scale uncertainties. 
Our results are essential input for applications of dispersive methods, including unitarity bounds and QCD sum rules.

\end{abstract}

\tableofcontents

\section{Introduction}

Flavour physics provides an ideal framework for indirect searches for physics beyond the Standard Model (SM).
This is accomplished by comparing precise experimental measurements with the corresponding SM predictions for observables in hadron decays, particularly in the decays of $B$ mesons.
In fact, the past three decades have been marked by a monumental effort to reduce both experimental and theoretical uncertainties, thereby enhancing the effectiveness of indirect searches.

The main obstacle to reducing theoretical uncertainties lies in the hadronic matrix elements --- such as decay constants and hadronic form factors --- whose predictions are governed by low-energy, and thus non-perturbative, QCD dynamics.
Techniques based on dispersion relations are extremely useful for predicting or constraining these matrix elements.
Indeed, dispersion relations --- one of the few exact results known in QCD --- relate matrix elements to perturbatively calculable quantities.
The two most commonly used techniques based on dispersion relations are unitarity bounds and QCD sum rules; see Refs.~\cite{Colangelo:2000dp,Caprini:2019osi,Gubernari:2020zil} for reviews.

The derivation of both unitarity bounds and QCD sum rules requires the calculation of (at least) a two-point correlator with quark currents.
Two-point correlators can be written as:
\begin{align}
    \label{eq:corrG}
    \Pi_\Gamma(q)
        & \equiv i\! \int\! d^4x\, e^{iq\cdot x} \bra{0} 
        T \left\{
            \bar{\psi}_1 \Gamma \psi_2 (x) \, \big(\bar{\psi}_1 \Gamma \psi_2(0)\big)^\dagger  
        \right\} \ket{0}
    \,,
\end{align}
where $\psi_{1,2}$ are quark fields and $\Gamma$ denotes a Dirac matrix or a product of Dirac matrices.
These correlators are typically calculated using an operator product expansion (OPE), which expresses the correlator as a sum over local operators. Each term consists of a Wilson coefficient, calculable perturbatively in the strong coupling constant $\alpha_s$, multiplied by a local operator of definite dimension, capturing non-perturbative effects. This naturally leads to a twofold expansion: in the operator dimensions and in $\alpha_s$.
In most cases, the dominant contribution to these OPEs comes from the dimension-zero operator, i.e., the identity operator, which requires the calculation of bubble diagrams (see, e.g., Ref.~\cite{Colangelo:2000dp} for a review).

Many results for such correlators are available in the literature. For (axial-)vector currents, namely $\Gamma = \gamma^\mu (\gamma_5)$, two-point correlators have been calculated at next-to-leading order (NLO) in $\alpha_s$ with two massive quarks and full momentum dependence~\cite{Djouadi:1993ss}.
These correlators have also been calculated at next-to-next-to-leading order (NNLO) with two massive quarks, but only at $q^2=0$ using both analytical and numerical methods~\cite{Grigo:2012ji}. The flavour-diagonal (axial-)vector correlators involving only massless quarks have been computed at five loops~\cite{Baikov:2001aa,Baikov:2008jh,Baikov:2010je,Baikov:2012er,Baikov:2012zm}.
For (pseudo)scalar currents, i.e.~$\Gamma = 1 (\gamma_5)$, two-point correlators have also been calculated at NLO in $\alpha_s$ with two massive quarks and full momentum dependence~\cite{Djouadi:1994gf}.
As for the (axial-)vector case, the (pseudo)scalar current correlators are known through NNLO with two massive quarks, but only at $q^2=0$~\cite{Grigo:2012ji}. For flavour-diagonal massless quarks, they are known at five loops \cite{Baikov:2001aa,Baikov:2005rw,Herzog:2017dtz}.
Results for the case where one of the quark masses is zero are known for both (pseudo)scalar and (axial-)vector currents at NNLO~\cite{Chetyrkin:2000mq,Chetyrkin:2001je,Chetyrkin:2006bj,Boughezal:2006xk,Hoff:2011ge,Maier:2011jd}.
For (axial-)tensor currents, namely $\Gamma = \sigma^{\mu\nu} (\gamma_5)$, two-point correlators have only been calculated at NLO with two massive quarks at $q^2=0$~\cite{Bharucha:2010im} or as a linear Taylor expansion in one of the masses~\cite{Pullin:2021ebn}.
This correlator has also been computed recently using lattice QCD~\cite{Harrison:2024iad}. For the massless flavour-diagonal case, the (axial-)tensor current correlator is known at NNLO~\cite{Chetyrkin:2010dx}.

In this article, we present the first complete analytical calculation of (axial-)tensor two-point correlators at NLO, with two massive quarks and full momentum dependence.
%
%
Furthermore, we estimate the scale uncertainties and compare our numerical results with those obtained in the lattice QCD study of Ref.~\cite{Harrison:2024iad}.

The remainder of this article is organised as follows.
In \refsec{calc}, we provide details of our calculation and present the analytical results.
In \refsec{num}, we give explicit numerical results for selected values of the quark masses and $q^2$.
Our conclusions are summarised in \refsec{conc}.

\section{Analytical calculation}
\label{sec:calc}

In this section, we first introduce our notation and review the one-loop contribution to the correlator (\refsubsec{onel}). 
We then present the two-loop calculation, providing explicit analytical results for the relevant correlators (\refsubsec{twol}). 
Finally, we compare our findings with existing results in the literature where they are available (\refsubsec{comp}).

\subsection{Theoretical framework and one-loop result}
\label{sec:onel}

We consider the correlators
\begin{align}
    \Pi_T^{\mu\nu}(q)
    & \equiv 
        i\! \int\! d^4x\, e^{iq\cdot x} \bra{0} 
        T \left\{
            \bar{\psi}_1 \sigma^{\mu\alpha}q_\alpha \psi_2 (x) \, \bar{\psi}_2 \sigma^{\nu\beta}q_\beta \psi_1 (0) 
        \right\} \ket{0}
    \nonumber\\ & =
    \label{eq:corrST}
    \left(\frac{q^\mu q^\nu}{q^2} - g^{\mu\nu}\right)
    \Pi_T(q^2)
    \,, \\
    \Pi_{AT}^{\mu\nu}(q)
    & \equiv 
        i\! \int\! d^4x\, e^{iq\cdot x} \bra{0} 
        T \left\{
            \bar{\psi}_1 \sigma^{\mu\alpha}\gamma_5 q_\alpha \psi_2 (x) \, \bar{\psi}_2 \sigma^{\nu\beta}\gamma_5 q_\beta \psi_1 (0) 
        \right\} \ket{0}
    \nonumber\\ & =
    \label{eq:corrSAT}
    \left(\frac{q^\mu q^\nu}{q^2} - g^{\mu\nu}\right)
    \Pi_{AT}(q^2)
    \,,
\end{align}
where $\psi_{1,2}$ are massive quark fields and $\sigma^{\mu\alpha} \equiv\frac{i}{2}[\gamma^\mu,\gamma^\alpha]$.
Note that the second line of \refeqa{corrST}{corrSAT} follows directly from Lorentz covariance.
Correlators such as $\Pi_T$ and $\Pi_{AT}$ satisfy the following identity, known as the \emph{dispersion relation} (see, e.g., Refs.~\cite{Colangelo:2000dp,Caprini:2019osi}):
\begin{align}
    \label{eq:disprel}
    \Pi_T(q^2) 
    & =
    \frac{1}{\pi} \int\limits_0^\infty d s \, \frac{\Im\,\Pi_T(s)}{s - q^2 - i\epsilon} 
    \,.
\end{align}
These correlators can be calculated using the OPE for $q^2
 \ll \left( m_1 + m_2 \right)^2  - \left( m_1 + m_2 \right) \Lambda_{\rm QCD} $, with $\Lambda_{\rm QCD}\sim 300\, \MeV$ (see, e.g., Refs.~\cite{Boyd:1997kz,Colangelo:2000dp}).
The leading power contribution to the OPE in the regime defined above can be calculated using standard perturbation theory.
In practice, it suffices to compute $\Pi_T$, since the results for $\Pi_{AT}$ can be obtained directly by performing the replacement $m_1 \to -m_1$~\cite{Djouadi:1993ss}.
For the remainder of this article, we therefore focus exclusively on $\Pi_T$, with the understanding that the corresponding results for $\Pi_{AT}$ can be easily derived.

\begin{figure}[t!]
    \centering

    \hspace*{1cm}
    \begin{subfigure}[b]{0.2\textwidth}
        \centering
        \includegraphics[width=\textwidth]{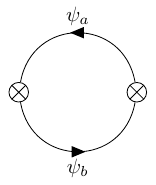}
        \caption{}
        \label{fig:1l}
    \end{subfigure}
    \hfill
    \begin{subfigure}[b]{0.2\textwidth}
        \centering
        \includegraphics[width=\textwidth]{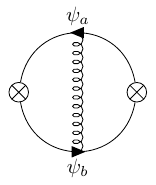}
        \caption{}
        \label{fig:2lA}
    \end{subfigure}
    \hfill
    \begin{subfigure}[b]{0.2\textwidth}
        \centering
        \includegraphics[width=\textwidth]{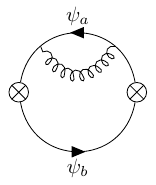}
        \caption{}
        \label{fig:2lB}
    \end{subfigure}
    \hfill
    \begin{subfigure}[b]{0.2\textwidth}
        \centering
        \includegraphics[width=\textwidth]{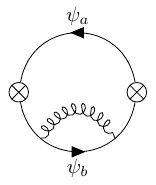}
        \caption{}
        \label{fig:2lC}
    \end{subfigure}
    \hspace*{1cm}

    \caption{Leading-power contribution to the OPE at LO (a) and NLO (b)-(d).
    The crossed circles denote the insertion points of the tensor current.}
    \label{fig:diags}
\end{figure}

The leading-power contribution to the OPE for $\Pi_T$ at leading order in $\alpha_s$ can be obtained straightforwardly, as it requires only the calculation of the one-loop diagram shown in \reffig{diags}(a).
Such one-loop calculations are fully automated using, for example, \texttt{Package~X}~\cite{Patel:2016fam} or \texttt{FeynCalc}~\cite{Shtabovenko:2023idz}.
This result reads
\begin{equation}
\begin{aligned} 
    \label{eq:1-loop}
    \Pi_T (q^2) \big|_{\text{1-loop}}
    & =
    \frac{q^2}{8\pi^2}   (6 m_1 m_2+q^2) \left(\frac{1}{\epsilon} + \log \left(\frac{\mu
   ^2}{m_1 m_2}\right)\right)
   \\* & +
    \frac{1}{48 \pi ^2 q^2}
    \Bigg(2 q^2 \left(-3 q^2 \left(m_1^2-12 m_1 m_2+m_2^2\right)-6
   \left(m_1^2-m_2^2\right)^2+8 (q^2)^2\right)
   \\* & +
   6 \left((m_1-m_2)^2-q^2\right) \left(2
   (m_1+m_2)^2+q^2\right) \lambda ^{1/2}\left(m_1^2,m_2^2,q^2\right) 
   \\* & \times
   \left(\log (2
   m_1 m_2)-\log \left(\lambda^{1/2}\left(m_1^2,m_2^2,q^2\right)+m_1^2+m_2^2-q^2\right)\right)
   \\* & +
   6
   (m_1-m_2) (m_1+m_2)^3 \left(2 (m_1-m_2)^2-3 q^2\right) \log
   \left(\frac{m_1}{m_2}\right)\Bigg)
   +
   \mathcal{O}(\epsilon)
   \,,
\end{aligned}
\end{equation} 
where  
$
    \lambda(m_1^2,m_2^2,q^2)
    \equiv
    \left(m_1^2 - m_2^2 - q^2\right)^2 - 4 m_2^2 \, q^2
$
is the Källén function. Note that throughout this work, we perform the substitution
\begin{equation}
    \mu^{2\epsilon}\to\frac{\mu^{2\epsilon}}{\Gamma(1+\epsilon)}\;,
\end{equation}
which allows us to renormalise the strong coupling in the \MSbar scheme by subtracting only pure poles in $\epsilon$.
The explicit result of \refeq{1-loop} is useful for illustrating how certain ultraviolet divergences are treated within the framework of dispersive techniques.
Typically, instead of renormalising vacuum-to-vacuum correlators such as $\Pi_T$, one subtracts from them the first few terms of their Taylor expansion in $q^2$.
This is equivalent to taking derivatives of \refeq{disprel} with respect to $q^2$:
\begin{align}
    \label{eq:subdisprel}
    \chi_T(Q^2;k) 
    \equiv
    \frac{1}{k!} \left[\frac{\partial}{\partial q^2}\right]^k
    \Pi_T(q^2) \Bigg|_{q^2=-Q^2} 
    & =
    \frac{1}{\pi} \int\limits_0^\infty \! d s \, \frac{\Im\,\Pi_T(s)}{(s + Q^2)^{k+1}} 
    \,.
\end{align}
Here, we define the quantity $\chi_T$ for later use.
It is straightforward to see that in this case at least three derivatives are required, and hence $k \geq 3$.
Note that beyond one-loop order, in addition to performing subtractions, the mass, the current, and the strong coupling must still be renormalised to cancel all divergences.

\subsection{Two-loop result}
\label{sec:twol}

We present the first calculation of the NLO corrections to the leading-power contribution in the OPE for $\Pi_T$ for two massive quarks and full $q^2$ dependence.
This involves evaluating the two-loop diagrams shown in \reffig{diags}(b)-(d).

In this study, the integration-by-parts (IBP) reduction can be performed relatively efficiently with contemporary computational tools, as the problem involves only three independent scales ($m_1$, $m_2$, and $q^2$) and the diagram topologies are comparatively simple, allowing for a manageable number of master integrals.
The IBP reduction is carried out using \texttt{FIRE}~\cite{Smirnov:2019qkx}, yielding a total of four master integrals not counting those which can be obtained by swapping the quark flavour labels. 
All of these master integrals have been evaluated analytically up to the required order in $\epsilon$ in the literature, with explicit expressions available in Ref.~\cite{Djouadi:1993ss}.
Combining these results yields the unrenormalised expression for $\Pi_T$.
For completeness we also obtain the expressions for the scalar $\Pi_S$ and vector $\Pi_{VL,VT}$ correlators, defined as
\begin{align}
    \Pi_S(q^2)
    & \equiv 
        i\! \int\! d^4x\, e^{iq\cdot x} \bra{0} 
        T \left\{
            \bar{\psi}_1 \psi_2 (x) \, \bar{\psi}_2  \psi_1 (0) 
        \right\} \ket{0}
    \,, \\
    \Pi_V^{\mu\nu}(q)
    & \equiv 
        i\! \int\! d^4x\, e^{iq\cdot x} \bra{0} 
        T \left\{
            \bar{\psi}_1 \gamma^\mu \psi_2 (x) \, \bar{\psi}_2 \gamma^\nu \psi_1 (0) 
        \right\} \ket{0}
    \nonumber\\ & =
    \frac{q^\mu q^\nu}{q^2}
    \Pi_{VL}(q^2)
    +
    \left(\frac{q^\mu q^\nu}{q^2} - g^{\mu\nu}\right) 
    \Pi_{VT}(q^2)
    \,.
\end{align}
We provide these results in machine-readable \texttt{Mathematica} format as ancillary files attached to the arXiv version of this paper.
For completeness, we also provide the LO results.
These files are named
\begin{align*}
     &\texttt{LO\_S.m} &
     &\texttt{LO\_VL.m} &
     &\texttt{LO\_VT.m} &
     &\texttt{LO\_TT.m} \\
     &\texttt{NLO\_S\_unren.m} &
     &\texttt{NLO\_VL\_unren.m} &
     &\texttt{NLO\_VT\_unren.m} &
     &\texttt{NLO\_TT\_unren.m}
\end{align*}

We now need to renormalise these expressions.
At two-loop order, one needs to renormalise the masses and the currents:
\begin{align}
    \label{eq:corrren}
    \Pi_T = \Pi_T \big|_{\rm unren.} + \Pi_T \big|_{\rm CT-mass} + \Pi_T \big|_{\rm CT-curr}
    \,,
\end{align}
where the counterterms are defined as 
\begin{align}
    \Pi_T \big|_{\rm CT-mass}
    & =
    - \delta m_1 \frac{\partial}{\partial m_1} 
    \Pi_T(q^2)\big|_{\text{1-loop}}
    - \delta m_2 \frac{\partial}{\partial m_2} 
    \Pi_T(q^2)\big|_{\text{1-loop}}
    \,,\\
    \Pi_T \big|_{\rm CT-curr}
    & =
    2 (Z_T-1) 
    \Pi_T(q^2)\big|_{\text{1-loop}}
    \,.
\end{align}
The renormalisation constants in the \MSbar scheme at one loop are given by\footnote{For the vector and scalar currents, the current renormalisation constants are given by $Z_V=1$ and $Z_S = 1 - 3\alpha_s/(4\pi\,\epsilon)C_F + {\cal O}{\left(\alpha_s^2\right)} $, respectively.}
\begin{align}
    \delta m_i 
    & \equiv 
    m_i (1-Z_m) 
    =
    m_i \frac{3\,\alpha_s}{4\,\pi \, \epsilon}
    C_F 
    + {\cal O}{\left(\alpha_s^2\right)} 
    \,,
    \\
    Z_T 
    & =
    1+\frac{\alpha_s}{4\,\pi\,\epsilon} C_F
    + {\cal O}{\left(\alpha_s^2\right)} 
    \,,
\end{align}
with $C_F =4/3$, while the mass renormalisation constant in the on-shell scheme at one loop is given by
\begin{equation}
\begin{aligned}
    \label{eq:massconsOS}
    \delta m_i 
    \equiv 
    m_i (1-Z_m)  
    & =
    \frac{m_i}{\epsilon}\frac{\alpha_s}{\pi}\bigg(\frac{\mu^2}{m_i^2}\bigg)^{\!\!\epsilon} C_F \frac{3-2\epsilon}{4-8\epsilon}+
    {\cal O}{\left(\alpha_s^2\right)} 
    \,.
\end{aligned}
\end{equation}
Therefore, using \refeqs{corrren}{massconsOS}, we obtain the renormalised expressions for the correlators.
As with the unrenormalised ones, these are provided in machine-readable \texttt{Mathematica} format as ancillary files attached to the arXiv version of this paper:
\begin{align*}
     &\texttt{NLO\_S\_MSbar.m} &
     &\texttt{NLO\_VL\_MSbar.m} &
     &\texttt{NLO\_VT\_MSbar.m} &
     &\texttt{NLO\_TT\_MSbar.m} \\
     &\texttt{NLO\_S\_OS.m} &
     &\texttt{NLO\_VL\_OS.m} &
     &\texttt{NLO\_VT\_OS.m} &
     &\texttt{NLO\_TT\_OS.m}
\end{align*}
For the reader’s convenience, we also provide the explicit expressions for the imaginary parts of these correlators, which can be extracted from the results above:
\begin{align*}
     &\texttt{Im\_LO\_S.m} &
     &\texttt{Im\_LO\_VL.m} &
     &\texttt{Im\_LO\_VT.m} &
     &\texttt{Im\_LO\_TT.m} \\
     &\texttt{Im\_NLO\_S\_MSbar.m} &
     &\texttt{Im\_NLO\_VL\_MSbar.m} &
     &\texttt{Im\_NLO\_VT\_MSbar.m} &
     &\texttt{Im\_NLO\_TT\_MSbar.m} \\
     &\texttt{Im\_NLO\_S\_OS.m} &
     &\texttt{Im\_NLO\_VL\_OS.m} &
     &\texttt{Im\_NLO\_VT\_OS.m} &
     &\texttt{Im\_NLO\_TT\_OS.m}
\end{align*}
Working with the imaginary part is often more convenient in the context of unitarity bounds, as taking derivatives --- i.e., performing subtractions --- then becomes trivial (see \refeq{subdisprel}).
Moreover, in the context of QCD sum rules, knowledge of the imaginary part constitutes a key ingredient~\cite{Colangelo:2000dp}.

The $\chi_\Gamma(Q^2;k)$ functions (i.e., the moments of the correlator) defined in \refeq{subdisprel} can in principle be obtained from the results above. However, since the lowest finite moments of the correlators at $Q^2=0$ are needed quite commonly, we also provide their explicit expressions in the files\footnote{
    It can be shown that the lowest admissible values of $k$ are $2$, $1$, $2$, and $3$ for $\chi_S$, $\chi_{VL}$, $\chi_{VT}$, and $\chi_T$, respectively.
}
\begin{align*}
     &\texttt{Moment\_LO\_S.m} &
     &\texttt{Moment\_NLO\_S\_unren.m} &
     &\texttt{Moment\_NLO\_S\_MSbar.m} &
     &\texttt{Moment\_NLO\_S\_OS.m} \\
     &\texttt{Moment\_LO\_VL.m} &
     &\texttt{Moment\_NLO\_VL\_unren.m} &
     &\texttt{Moment\_NLO\_VL\_MSbar.m} &
     &\texttt{Moment\_NLO\_VL\_OS.m} \\
     &\texttt{Moment\_LO\_VT.m} &
     &\texttt{Moment\_NLO\_VT\_unren.m} &
     &\texttt{Moment\_NLO\_VT\_MSbar.m} &
     &\texttt{Moment\_NLO\_VT\_OS.m} \\
     &\texttt{Moment\_LO\_TT.m} &
     &\texttt{Moment\_NLO\_TT\_unren.m} &
     &\texttt{Moment\_NLO\_TT\_MSbar.m} &
     &\texttt{Moment\_NLO\_TT\_OS.m}
\end{align*}

\subsection{Comparison with previous results}
\label{sec:comp}

We conclude this section by comparing our results with the literature.
As a validation of our calculation, we compare our results for $\Pi_S$ and $\Pi_V$ with those reported in Refs.~\cite{Djouadi:1993ss,Grigo:2012ji} and find exact agreement.
We compare our results with Ref.~\cite{Pullin:2021ebn}, which presents the NLO expression for $\Im\,\Pi_T$ in the regime $m_1 \ll m_2$, including terms up to ${\cal O}(m_1/m_2)$, and find exact agreement.

Finally, we verify our results against those of Ref.~\cite{Bharucha:2010im}, which presents $\Pi_S$, $\Pi_V$, and $\Pi_T$ at $q^2=0$ with the minimal number of subtractions (i.e.~$\chi_S(0;2)$, $\chi_{VL}(0;1)$, $\chi_{VT}(0;2)$, and $\chi_T(0;3)$).
While Ref.~\cite{Bharucha:2010im} does not specify the treatment of current renormalisation, we find agreement with their results only if the last term in \refeq{corrren} is omitted. 
When this term is included instead, the expression for $\Pi_V$ remains unchanged, since the vector current does not require renormalisation, whereas our corresponding results for the $\mathcal{O}(\epsilon^0)$ parts of $\Pi_S$ and $\Pi_T$ are different and disagree with the results of Ref.~\cite{Bharucha:2010im}. As expected, we find that the moments and the imaginary parts of $\Pi_S$ and $\Pi_T$ are free of $\epsilon$-poles only when the current renormalisation is performed.

Therefore, our result for $\Pi_T$ not only supersedes previous literature by providing the full $m_1$, $m_2$, and $q^2$ dependence, but also resolves an inconsistency of earlier results.

\section{Numerical results}
\label{sec:num}

In this section, we provide explicit numerical results based on the analytical expressions derived in the previous section.
\reftab{numres} reports numerical values of the functions $\chi_T(Q^2;k)$ and $\chi_{AT}(Q^2;k)$ defined in \refeq{subdisprel}.
The values at $Q^2 = 0$ and $k = 3$ are of particular importance, as they are most commonly used in the context of unitarity bounds~\cite{Bharucha:2010im,Gubernari:2023puw,Bordone:2025jur}.
Nevertheless, other values of $Q^2$ and $k$ can also be relevant in that context~\cite{Boyd:1997kz,Gopal:2024mgb}.

We adopt the \MSbar scheme for the quark masses, since it provides a short-distance mass definition that avoids the infrared sensitivities inherent in the on-shell scheme. 
In particular, the pole mass suffers from renormalon ambiguities of order $\Lambda_{\rm QCD}$, making it ill-suited for perturbative analyses of correlators such as those considered in this work~\cite{Egner:2024lay}.\footnote{
    See  also e.g.~Ref.~\cite{Beneke:2021lkq} for a review of quark-mass renormalisation schemes. In particular, it is argued in Refs.~\cite{Egner:2024lay,Beneke:2021lkq} that the kinetic mass scheme is the most appropriate for semi-leptonic decays of heavy-light mesons. However, since we are performing a calculation at a relatively low order in perturbation theory, such considerations likely lie beyond the precision attainable here. In particular, we observe good apparent perturbative convergence through NLO using the $\overline{\text{MS}}$ scheme, which a posteriori justifies this choice.}
The values of the quark masses used in our numerical analysis are also listed in \reftab{numres}.
The central values are taken from the Particle Data Group~\cite{PDG}, while their uncertainties are neglected, since their effect on the final results is much smaller than that of the scale uncertainties.
We then evolve the quark masses to the required scale using the three-loop running implemented in \texttt{RunDec}~\cite{Chetyrkin:2000yt,Herren:2017osy}, with~$\alpha_s(m_Z)=0.1180$~\cite{PDG}.

\begin{table}[t!]
    \newcommand{\pp}{\phantom{+}}
    \centering
    \renewcommand{\arraystretch}{1.5}
    \resizebox{1.0\textwidth}{!}{%
    \begin{tabular}{ccc}
        \toprule
        Value of the masses                         &
        $\chi_\Gamma(Q^2;k)$                        &
        Central value $\pm$ unc. [$\GeV^{-2}$]         \\
        \toprule
        \multirow{4}{*}{\makecell{
            $m_1=0$ \\ 
            $m_2=\overline{m}_b(\overline{m}_b)=4.183\,\GeV$
        }}                                          &
        \multirow{2}{*}{\makecell{
            $\chi_T(0;3)=\chi_{AT}(0;3)$             
        }}                                          & 
        \multirow{2}{*}{\makecell{
            $(4.060_{-0.589}^{+0.119})\cdot 10^{-4}$    
        }}                                          \\
                                                    \\
                                                    &
        \multirow{2}{*}{\makecell{
            $\chi_T(Q_b^2;3)=\chi_{AT}(Q_b^2;3)$ 
        }}                                          & 
        \multirow{2}{*}{\makecell{
            $(1.888_{-0.175}^{+0.081})\cdot 10^{-4}$    
        }}                                          \\
                                                    \\
        \midrule 
        \multirow{4}{*}{\makecell{
            $m_1=\overline{m}_s(2\,\GeV)=0.0935\,\GeV$ \\ 
            $m_2=\overline{m}_b(\overline{m}_b)=4.183\,\GeV$
        }}                                          &
        $\chi_T(0;3)$                               & 
        $(4.235_{-0.631}^{+0.140})\cdot 10^{-4}$    \\
                                                    &
        $\chi_{AT}(0;3)$                            & 
        $(3.877_{-0.557}^{+0.108})\cdot 10^{-4}$    \\
                                                    &
        $\chi_T(Q_b^2;3)$                           & 
        $(1.934_{-0.180}^{+0.086})\cdot 10^{-4}$    \\
                                                    &
        $\chi_{AT}(Q_b^2;3)$                        & 
        $(1.842_{-0.172}^{+0.078})\cdot 10^{-4}$    \\
        \midrule 
        \multirow{4}{*}{\makecell{ 
            $m_1=\overline{m}_c(\overline{m}_c)=1.273\,\GeV$\\ 
            $m_2=\overline{m}_b(\overline{m}_b)=4.183\,\GeV$
        }}                                          &
        $\chi_T(0;3)$                               & 
        $(4.857_{-0.723}^{+0.173})\cdot 10^{-4}$    \\
                                                    &
        $\chi_{AT}(0;3)$                            & 
        $(2.400_{-0.335}^{+0.078})\cdot 10^{-4}$    \\
                                                    &
        $\chi_T(Q_b^2;3)$                           & 
        $(2.248_{-0.215}^{+0.114})\cdot 10^{-4}$    \\
                                                    &
        $\chi_{AT}(Q_b^2;3)$                        & 
        $(1.404_{-0.149}^{+0.054})\cdot 10^{-4}$    \\
        \bottomrule
    \end{tabular}
    }
    \caption{
      Central values (with all scales fixed to $\mu_b \equiv \overline{m}_b(\overline{m}_b) = 4.183\,\GeV$) and corresponding uncertainties of the functions $\chi_T(Q^2;k)$ and $\chi_{AT}(Q^2;k)$ for $Q^2=\{0,Q_b^2\equiv\overline{m}_b^2(\overline{m}_b)\}$ and different quark masses. 
      See the text for further details.
    \label{tab:numres} 
    }
\end{table}

We choose as reference scale $\mu_b \equiv \overline{m}_b(\overline{m}_b) = 4.183\,\GeV$.
The theoretical uncertainties are estimated by independently varying the four scales that enter our calculation: two associated with the quark masses, one with the strong coupling, and one with the current coupling. 
For each scale, the correlators are evaluated at three points, $\{\mu_b/2, \mu_b, 2\mu_b\}$, discarding extreme configurations in which two scales differ by more than a factor of two.
This restricted independent variation results in $31$ distinct scale combinations.
This prescription is a straightforward extension of the usual 7-point scale variation adopted for the case of two arbitrary scales to the case of four scales. There are several examples in the literature (see e.g.~Refs.~\cite{Catani:2020tko,Garzelli:2020fmd,Czakon:2021ohs,Chen:2022gpk,Duhr:2022yyp,Isaacson:2022rts}) where a similar extension to three scales was employed and found to yield good estimates of the scale uncertainties.
The correlator currents are then consistently evolved to the reference scale $\mu_b$ using the renormalisation group equation
\begin{equation}\label{eq:PiRGE}
    \frac{d\Pi_T}{d\ln{\mu^2}} 
    =2\,\gamma_T\,\Pi_T\,,
\end{equation}
where $\gamma_T$ is the anomalous dimension of the tensor current~\cite{Broadhurst:1994se}:
\begin{equation}
\begin{aligned}
    \frac{d(\bar{\psi}_1 \sigma^{\mu\alpha}q_\alpha \psi_2)}{d\ln{\mu^2}} &=  \gamma_T\,\bar{\psi}_1 \sigma^{\mu\alpha}q_\alpha \psi_2\,,& 
    \gamma_T &= \sum_{i=0}^\infty\gamma_{T,i}\bigg(\frac{\alpha_s}{\pi}\bigg)^{\!\!i+1}, \\
    \gamma_{T,0} &= -\frac{1}{3}\,,& 
    \gamma_{T,1} &= -\frac{181}{72} + n_f\frac{13}{108}\,.&
\end{aligned}
\end{equation}
The solution of \refeq{PiRGE} is given by
\begin{equation}
\begin{aligned}
    \Pi_T(q^2;\mu_b) 
    & =
    \Pi_T(q^2;\mu_0) 
    \exp\Bigg\{
        \int\limits_{\alpha_s(\mu_0)}^{\alpha_s(\mu_b)} \frac{2\gamma_T(\alpha)}{\beta(\alpha)} \, d\alpha
    \Bigg\}  \,,
\end{aligned}
\end{equation}
where~\cite{Caswell:1974gg,Jones:1974mm}
\begin{align}
   \frac{d\alpha_s}{d\ln{\mu^2}} &= \:\beta\;,& 
   \beta &= -\alpha_s\sum_{i=0}^\infty\beta_{i}\bigg(\frac{\alpha_s}{4\pi}\bigg)^{i+1} , &
    \beta_0 &= 11 - \frac{2}{3} n_f , &
    \beta_1 &= 102 - \frac{38}{3} n_f , &
\end{align}
is the QCD beta function.
This procedure ensures that the dominant scale dependence, already captured at NLO, largely cancels upon running, while the residual dependence provides a reliable estimate of the missing higher-order corrections.

We remark that the uncertainties on $\chi_T(Q^2;k)$ are sizeable, with the lower uncertainty reaching up to $\sim 15\%$ and the upper uncertainty up to $\sim 5\%$. 
Nevertheless, these uncertainties are usually neglected in the context of unitarity bounds.
A straightforward way to incorporate them is to adopt the upper values of $\chi_T(Q^2;k)$, which constitutes the most conservative choice (see, e.g., Refs.~\cite{Boyd:1997kz,Caprini:1997mu}).
The magnitude of these uncertainties further motivates the necessity of an NNLO calculation of $\chi_T(Q^2;k)$.
\\

\begin{figure}[t!]
    \centering

    \begin{subfigure}{0.48\textwidth}
        \centering
        \includegraphics[width=\textwidth]{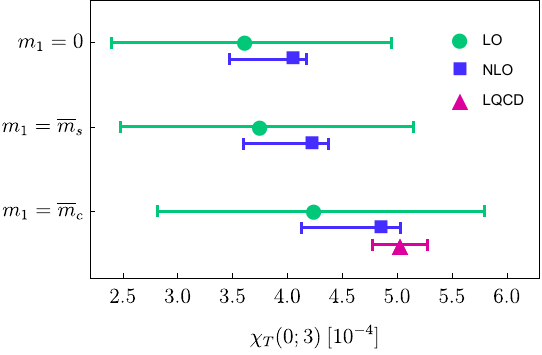}
    \end{subfigure}
    \hfill
    \begin{subfigure}{0.48\textwidth}
        \centering
        \includegraphics[width=\textwidth]{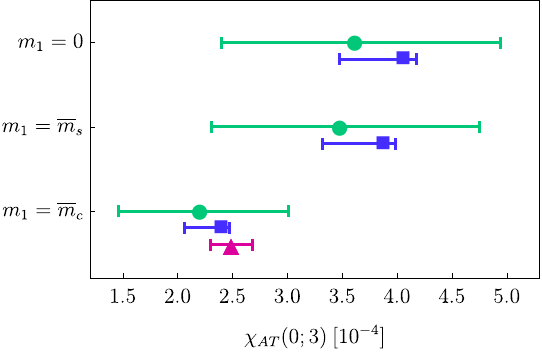}
    \end{subfigure}

    \caption{Comparison of our LO (green) and NLO (blue) results with lattice QCD data\cite{Harrison:2024iad} (magenta) for different values of $m_1$.
    The central values are obtained with all scales fixed to $\mu_b \equiv \overline{m}_b(\overline{m}_b) = 4.183\,\GeV$, whereas the procedure used to determine the associated uncertainties is described in the text.
    }
    \label{fig:chiTAT}
\end{figure}

As discussed in the previous section, the results of Ref.~\cite{Bharucha:2010im} apparently omit the current renormalisation.
This induces a shift of the central values by about $5$–$7\%$, which, although not very large, is nevertheless non-negligible.

In \reffig{chiTAT} we show our LO and NLO results for $\chi_T(0;3)$ and $\chi_{AT}(0;3)$. 
As seen in the figure, the inclusion of NLO corrections significantly reduces the uncertainties compared to LO.
For the charm-quark case, we also compare with lattice QCD results of Ref.~\cite{Harrison:2024iad}, which are consistent within the uncertainties. 
Note that our perturbative calculation and the lattice-QCD results can, to a good approximation, be regarded as two independent determinations, and hence their results can be combined to achieve smaller uncertainties:
\begin{align}
    \chi_T(0;3) & = (4.92 \pm 0.22)\cdot 10^{-4}\,\GeV^{-2} \,,&
    \chi_{AT}(0;3) & = (2.39 \pm 0.14)\cdot 10^{-4}\,\GeV^{-2} \,.&
\end{align}
We emphasise, however, that a non-trivial correlation exists due to the common dependence on the quark masses. 
Our rationale is that the uncertainties associated with the quark masses are essentially negligible compared to the scale uncertainties in our calculation and to the discretisation effects present in Ref.~\cite{Harrison:2024iad}, such that treating the two determinations as effectively independent is justified. 
To obtain the quoted averages, the central values of our results were shifted to the means of their asymmetric distributions, thereby rendering the uncertainties symmetric.

\begin{figure}[t!]
    \centering

    \begin{subfigure}{0.48\textwidth}
        \centering
        \includegraphics[width=\textwidth]{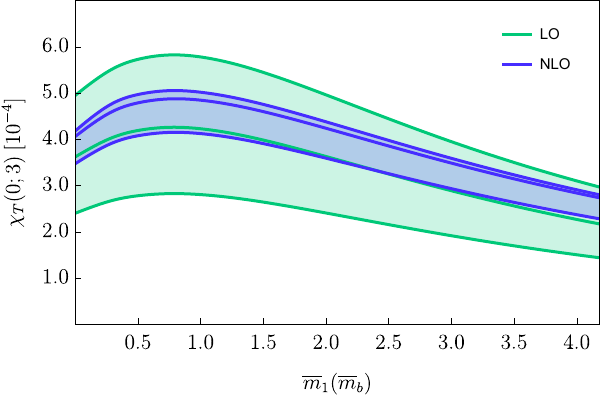}
    \end{subfigure}
    \hfill
    \begin{subfigure}{0.48\textwidth}
        \centering
        \includegraphics[width=\textwidth]{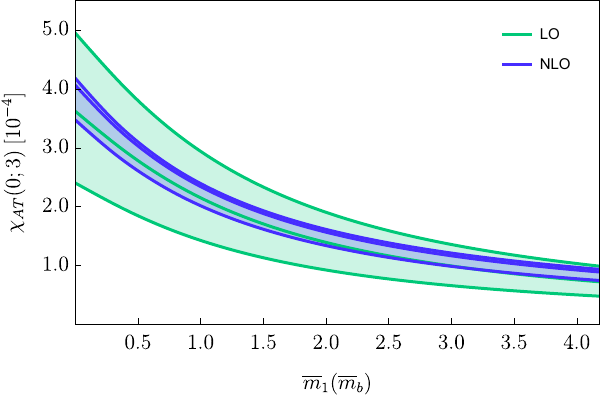}
    \end{subfigure}

    \caption{$\chi_T(0;3)$ (left) and $\chi_{AT}(0;3)$ (right) as functions of $\overline{m}_1(\overline{m}_b)$. The LO results are shown in green, while the NLO results are shown in blue.
    The central values are obtained with all scales fixed to $\mu_b \equiv \overline{m}_b(\overline{m}_b) = 4.183\,\GeV$, whereas the procedure used to determine the associated uncertainties is described in the text.
    }
    \label{fig:chiTAT_massdep}
\end{figure}

In order to further elucidate the behaviour of the mass dependence of $\chi_{T}(0;3)$ and $\chi_{AT}(0;3)$, we display in \reffig{chiTAT_massdep} their values as functions of $\overline{m}_1(\overline{m}_b)$. 
It is straightforward to observe that the perturbative convergence pattern remains stable over the full range of masses scanned, i.e., the separation between LO and NLO predictions does not increase when the mass is varied.
This behaviour supports the interpretation that the perturbative series remains well controlled across the interval considered, reinforcing the robustness of the quoted central values and their associated uncertainties. 
Nevertheless, an NNLO determination would be highly desirable in order to further consolidate these findings.

\section{Summary and conclusion}
\label{sec:conc}

We have presented the first complete analytical calculation of the vacuum-to-vacuum correlator of two flavour-non-diagonal tensor currents of massive quarks, retaining the full dependence on the two quark masses and the external momentum. Our results extend previous studies by providing NLO corrections for arbitrary $q^2$, thereby filling a gap in the literature and correcting an inconsistency in the results of Ref.~\cite{Bharucha:2010im}, which seemingly omitted the renormalisation of the tensor current. 

We provide all analytical expressions in machine-readable \texttt{Mathematica} format, facilitating their use in applications such as unitarity bounds and QCD sum rules. Using these expressions, we have presented numerical results for selected quark-mass values and demonstrated that the NLO corrections reduce the theoretical uncertainties compared to the LO predictions. The uncertainties, though reduced, remain non-negligible and should be accounted for in precision applications. 

Our results show good agreement with a recent lattice QCD calculation~\cite{Harrison:2024iad}, providing an independent cross-check and enabling a combination of perturbative and non-perturbative determinations to further improve precision. Finally, the magnitude of the remaining uncertainties motivates the calculation of NNLO corrections, which would be the natural next step towards fully controlled predictions for tensor-current correlators.  

Overall, the work presented here provides essential theoretical input for dispersive analyses in flavour physics and represents a significant step forward in the precise determination of hadronic matrix elements involving tensor currents.

\section*{Acknowledgements}

We thank the Cambridge Pheno Working Group for insightful discussions.
NG is grateful to D. van Dyk for valuable input on the ideas underlying this work.
NG thanks the authors of Ref.~\cite{Bharucha:2010im} for providing their results in a machine-readable form and for helpful discussion.

\sloppy This work has been partially supported by STFC consolidated grants ST/T000694/1 and ST/X000664/1.
EL receives further support from the Alfred Kordelin Foundation.

\bibliographystyle{JHEP}
\bibliography{references}

@article{Grigo:2012ji,
    author = "Grigo, Jonathan and Hoff, Jens and Marquard, Peter and Steinhauser, Matthias",
    title = "{Moments of heavy quark correlators with two masses: exact mass dependence to three loops}",
    eprint = "1206.3418",
    archivePrefix = "arXiv",
    primaryClass = "hep-ph",
    reportNumber = "SFB-CPP-12-39, TTP12-20",
    doi = "10.1016/j.nuclphysb.2012.07.007",
    journal = "Nucl. Phys. B",
    volume = "864",
    pages = "580--596",
    year = "2012"
}

@article{Bharucha:2010im,
    author = "Bharucha, Aoife and Feldmann, Thorsten and Wick, Michael",
    title = "{Theoretical and Phenomenological Constraints on Form Factors for Radiative and Semi-Leptonic B-Meson Decays}",
    eprint = "1004.3249",
    archivePrefix = "arXiv",
    primaryClass = "hep-ph",
    reportNumber = "IPPP-10-31, DCPT-10-62, TUM-HEP-756-10",
    doi = "10.1007/JHEP09(2010)090",
    journal = "JHEP",
    volume = "09",
    pages = "090",
    year = "2010"
}

@article{Boyd:1997kz,
    author = "Boyd, C. Glenn and Grinstein, Benjamin and Lebed, Richard F.",
    title = "{Precision corrections to dispersive bounds on form-factors}",
    eprint = "hep-ph/9705252",
    archivePrefix = "arXiv",
    reportNumber = "CMU-HEP-97-07A, UCSD-PTH-97-12",
    doi = "10.1103/PhysRevD.56.6895",
    journal = "Phys. Rev. D",
    volume = "56",
    pages = "6895--6911",
    year = "1997"
}

@article{Caprini:1997mu,
    author = "Caprini, Irinel and Lellouch, Laurent and Neubert, Matthias",
    title = "{Dispersive bounds on the shape of $\bar{B} \to D^{(*)}\ell \bar \nu$ form-factors}",
    eprint = "hep-ph/9712417",
    archivePrefix = "arXiv",
    reportNumber = "CERN-TH-97-091, CPT-97-P3480",
    doi = "10.1016/S0550-3213(98)00350-2",
    journal = "Nucl. Phys. B",
    volume = "530",
    pages = "153--181",
    year = "1998"
}

@phdthesis{Gubernari:2020zil,
    author = "Gubernari, Nico",
    title = "{Applications of Light-Cone Sum Rules in Flavour Physics}",
    school = "Munich, Tech. U.",
    year = "2020"
}

@article{Colangelo:2000dp,
    author = "Colangelo, Pietro and Khodjamirian, Alexander",
    editor = "Shifman, M. and Ioffe, Boris",
    title = "{QCD sum rules, a modern perspective}",
    eprint = "hep-ph/0010175",
    archivePrefix = "arXiv",
    reportNumber = "CERN-TH-2000-296, BARI-TH-2000-394",
    doi = "10.1142/9789812810458_0033",
    pages = "1495--1576",
    month = "10",
    year = "2000"
}

@article{Harrison:2024iad,
    author = "Harrison, Judd",
    title = "{$\bar{b}c$ susceptibilities from fully relativistic lattice QCD}",
    eprint = "2405.01390",
    archivePrefix = "arXiv",
    primaryClass = "hep-lat",
    doi = "10.1103/PhysRevD.110.054506",
    journal = "Phys. Rev. D",
    volume = "110",
    number = "5",
    pages = "054506",
    year = "2024"
}

@book{Caprini:2019osi,
    author = "Caprini, Irinel",
    title = "{Functional Analysis and Optimization Methods in Hadron Physics}",
    doi = "10.1007/978-3-030-18948-8",
    isbn = "978-3-030-18947-1, 978-3-030-18948-8",
    publisher = "Springer",
    series = "SpringerBriefs in Physics",
    year = "2019"
}

@article{Djouadi:1993ss,
    author = "Djouadi, A. and Gambino, P.",
    title = "{Electroweak gauge bosons selfenergies: Complete QCD corrections}",
    eprint = "hep-ph/9309298",
    archivePrefix = "arXiv",
    reportNumber = "UDEM-LPN-TH-93-169, NYU-TH-93-09-04",
    doi = "10.1103/PhysRevD.49.3499",
    journal = "Phys. Rev. D",
    volume = "49",
    pages = "3499--3511",
    year = "1994",
    note = "[Erratum: Phys.Rev.D 53, 4111 (1996)]"
}

@article{Shtabovenko:2023idz,
    author = "Shtabovenko, Vladyslav and Mertig, Rolf and Orellana, Frederik",
    title = "{FeynCalc 10: Do multiloop integrals dream of computer codes?}",
    eprint = "2312.14089",
    archivePrefix = "arXiv",
    primaryClass = "hep-ph",
    reportNumber = "P3H-23-089, TTP23-056, SI-HEP-2023-27",
    doi = "10.1016/j.cpc.2024.109357",
    journal = "Comput. Phys. Commun.",
    volume = "306",
    pages = "109357",
    year = "2025"
}

@article{Patel:2016fam,
    author = "Patel, Hiren H.",
    title = "{Package-X 2.0: A Mathematica package for the analytic calculation of one-loop integrals}",
    eprint = "1612.00009",
    archivePrefix = "arXiv",
    primaryClass = "hep-ph",
    doi = "10.1016/j.cpc.2017.04.015",
    journal = "Comput. Phys. Commun.",
    volume = "218",
    pages = "66--70",
    year = "2017"
}

@article{Smirnov:2019qkx,
    author = "Smirnov, A. V. and Chukharev, F. S.",
    title = "{FIRE6: Feynman Integral REduction with modular arithmetic}",
    eprint = "1901.07808",
    archivePrefix = "arXiv",
    primaryClass = "hep-ph",
    reportNumber = "TTP19-006",
    doi = "10.1016/j.cpc.2019.106877",
    journal = "Comput. Phys. Commun.",
    volume = "247 ",
    pages = "106877",
    year = "2020"
}

@article{Pullin:2021ebn,
    author = "Pullin, Ben and Zwicky, Roman",
    title = "{Radiative decays of heavy-light mesons and the $ {f}_{H,{H}^{\ast },{H}_1}^{(T)} $ decay constants}",
    eprint = "2106.13617",
    archivePrefix = "arXiv",
    primaryClass = "hep-ph",
    reportNumber = "CP3-Origins-2020-13 DNRF90",
    doi = "10.1007/JHEP09(2021)023",
    journal = "JHEP",
    volume = "09",
    pages = "023",
    year = "2021"
}

@article{Gopal:2024mgb,
    author = "Gopal, Abinand and Gubernari, Nico",
    title = "{Unitarity bounds with subthreshold and anomalous cuts for b-hadron decays}",
    eprint = "2412.04388",
    archivePrefix = "arXiv",
    primaryClass = "hep-ph",
    doi = "10.1103/PhysRevD.111.L031501",
    journal = "Phys. Rev. D",
    volume = "111",
    number = "3",
    pages = "L031501",
    year = "2025"
}

@article{Bordone:2025jur,
    author = "Bordone, Marzia and Gubernari, Nico and Jung, Martin and van Dyk, Danny",
    title = "{Challenging $ {\overline{B}}_{(s)}\to {D}_{(s)}^{\left(\ast \right)} $ form factors with the heavy quark expansion}",
    eprint = "2507.03569",
    archivePrefix = "arXiv",
    primaryClass = "hep-ph",
    reportNumber = "CERN-TH-2025-092, EOS-2025-03, IPPP/25/25, P3H-25-032, SI-HEP-2025-10, ZU-TH 35/25",
    doi = "10.1007/JHEP11(2025)051",
    journal = "JHEP",
    volume = "11",
    pages = "051",
    year = "2025"
}

@article{Gubernari:2023puw,
    author = "Gubernari, Nico and Reboud, M{\'e}ril and van Dyk, Danny and Virto, Javier",
    title = "{Dispersive analysis of $B \to K^{(*)}$ and $B_{s} \to \phi$ form factors}",
    eprint = "2305.06301",
    archivePrefix = "arXiv",
    primaryClass = "hep-ph",
    reportNumber = "EOS-2023-02, IPPP/23/22, P3H-23-026, SI-HEP-2023-09",
    doi = "10.1007/JHEP12(2023)153",
    journal = "JHEP",
    volume = "12",
    pages = "153",
    year = "2023",
    note = "[Erratum: JHEP 01, 125 (2025)]"
}

@article{Egner:2024lay,
    author = {Egner, Manuel and Fael, Matteo and Lenz, Alexander and Piscopo, Maria Laura and Rusov, Aleksey V. and Sch{\"o}nwald, Kay and Steinhauser, Matthias},
    title = "{Total decay rates of B mesons at NNLO-QCD}",
    eprint = "2412.14035",
    archivePrefix = "arXiv",
    primaryClass = "hep-ph",
    reportNumber = "TUM-HEP-1545/24, P3H-24-101, SI-HEP-2024-31, TTP24-046, Nikhef
  2024-019, ZU-TH 67/24",
    doi = "10.1007/JHEP04(2025)106",
    journal = "JHEP",
    volume = "04",
    pages = "106",
    year = "2025"
}

@article{PDG,
    author = "Navas, S. and others",
    collaboration = "Particle Data Group",
    title = "{Review of particle physics}",
    doi = "10.1103/PhysRevD.110.030001",
    journal = "Phys. Rev. D",
    volume = "110",
    number = "3",
    pages = "030001",
    year = "2024"
}

@article{Chetyrkin:2000yt,
    author = "Chetyrkin, K. G. and Kuhn, Johann H. and Steinhauser, M.",
    title = "{RunDec: A Mathematica package for running and decoupling of the strong coupling and quark masses}",
    eprint = "hep-ph/0004189",
    archivePrefix = "arXiv",
    reportNumber = "DESY-00-034, TTP-00-05",
    doi = "10.1016/S0010-4655(00)00155-7",
    journal = "Comput. Phys. Commun.",
    volume = "133",
    pages = "43--65",
    year = "2000"
}

@article{Broadhurst:1994se,
    author = "Broadhurst, David J. and Grozin, A. G.",
    title = "{Matching QCD and heavy-quark effective theory heavy-light currents at two loops and beyond}",
    eprint = "hep-ph/9410240",
    archivePrefix = "arXiv",
    reportNumber = "OUT-4102-52",
    doi = "10.1103/PhysRevD.52.4082",
    journal = "Phys. Rev. D",
    volume = "52",
    pages = "4082--4098",
    year = "1995"
}

@article{Caswell:1974gg,
    author = "Caswell, William E.",
    title = "{Asymptotic Behavior of Nonabelian Gauge Theories to Two Loop Order}",
    reportNumber = "PRINT-74-1058 (PRINCETON)",
    doi = "10.1103/PhysRevLett.33.244",
    journal = "Phys. Rev. Lett.",
    volume = "33",
    pages = "244",
    year = "1974"
}

@article{Jones:1974mm,
    author = "Jones, D. R. T.",
    title = "{Two Loop Diagrams in Yang-Mills Theory}",
    reportNumber = "OXFORD-TP-10-74",
    doi = "10.1016/0550-3213(74)90093-5",
    journal = "Nucl. Phys. B",
    volume = "75",
    pages = "531",
    year = "1974"
}

@article{Beneke:2021lkq,
    author = "Beneke, Martin",
    title = "{Pole mass renormalon and its ramifications}",
    eprint = "2108.04861",
    archivePrefix = "arXiv",
    primaryClass = "hep-ph",
    reportNumber = "TUM-HEP-1358/21",
    doi = "10.1140/epjs/s11734-021-00268-w",
    journal = "Eur. Phys. J. ST",
    volume = "230",
    number = "12-13",
    pages = "2565--2579",
    year = "2021"
}

@article{Catani:2020tko,
    author = "Catani, Stefano and Devoto, Simone and Grazzini, Massimiliano and Kallweit, Stefan and Mazzitelli, Javier",
    title = "{Top-quark pair hadroproduction at NNLO: differential predictions with the $\overline{MS}$ mass}",
    eprint = "2005.00557",
    archivePrefix = "arXiv",
    primaryClass = "hep-ph",
    reportNumber = "ZU-TH 10/20, MPP-2020-52",
    doi = "10.1007/JHEP08(2020)027",
    journal = "JHEP",
    volume = "08",
    number = "08",
    pages = "027",
    year = "2020"
}

@article{Garzelli:2020fmd,
    author = "Garzelli, M. V. and Kemmler, L. and Moch, S. and Zenaiev, O.",
    title = "{Heavy-flavor hadro-production with heavy-quark masses renormalized in the ${\overline{\rm MS}}$, MSR and on-shell schemes}",
    eprint = "2009.07763",
    archivePrefix = "arXiv",
    primaryClass = "hep-ph",
    reportNumber = "DESY-20-151, DESY 20-151",
    doi = "10.1007/JHEP04(2021)043",
    journal = "JHEP",
    volume = "04",
    pages = "043",
    year = "2021"
}

@article{Czakon:2021ohs,
    author = "Czakon, Micha\l{} and Generet, Terry and Mitov, Alexander and Poncelet, Rene",
    title = "{B-hadron production in NNLO QCD: application to LHC t$ \overline{t} $ events with leptonic decays}",
    eprint = "2102.08267",
    archivePrefix = "arXiv",
    primaryClass = "hep-ph",
    reportNumber = "TTK-21-05, P3H-21-011, Cavendish-HEP-21/02",
    doi = "10.1007/JHEP10(2021)216",
    journal = "JHEP",
    volume = "10",
    pages = "216",
    year = "2021"
}

@article{Chen:2022gpk,
    author = {Chen, X. and Gehrmann, T. and Glover, E. W. N. and H{\"o}fer, M. and Huss, A. and Sch{\"u}rmann, R.},
    title = "{Single photon production at hadron colliders at NNLO QCD with realistic photon isolation}",
    eprint = "2205.01516",
    archivePrefix = "arXiv",
    primaryClass = "hep-ph",
    reportNumber = "ZU-TH 16/22, KA-TP-12-2022, P3H-22-044, IPPP/22/30, IPPP/22/30,
  CERN-TH-2022-072, LMU-ASC 17/22, CERN-TH-2022-072",
    doi = "10.1007/JHEP08(2022)094",
    journal = "JHEP",
    volume = "08",
    pages = "094",
    year = "2022"
}

@article{Duhr:2022yyp,
    author = "Duhr, Claude and Mistlberger, Bernhard and Vita, Gherardo",
    title = "{Four-Loop Rapidity Anomalous Dimension and Event Shapes to Fourth Logarithmic Order}",
    eprint = "2205.02242",
    archivePrefix = "arXiv",
    primaryClass = "hep-ph",
    reportNumber = "BONN-TH-2022-11, SLAC-PUB-17675",
    doi = "10.1103/PhysRevLett.129.162001",
    journal = "Phys. Rev. Lett.",
    volume = "129",
    number = "16",
    pages = "162001",
    year = "2022"
}

@article{Isaacson:2022rts,
    author = "Isaacson, Joshua and Fu, Yao and Yuan, C. -P.",
    title = "{resbos2 and the CDF W mass measurement}",
    eprint = "2205.02788",
    archivePrefix = "arXiv",
    primaryClass = "hep-ph",
    reportNumber = "FERMILAB-PUB-22-374-T, MSUHEP-22-017",
    doi = "10.1103/PhysRevD.110.094023",
    journal = "Phys. Rev. D",
    volume = "110",
    number = "9",
    pages = "094023",
    year = "2024"
}

@article{Herren:2017osy,
    author = "Herren, Florian and Steinhauser, Matthias",
    title = "{Version 3 of RunDec and CRunDec}",
    eprint = "1703.03751",
    archivePrefix = "arXiv",
    primaryClass = "hep-ph",
    reportNumber = "TTP17-011",
    doi = "10.1016/j.cpc.2017.11.014",
    journal = "Comput. Phys. Commun.",
    volume = "224",
    pages = "333--345",
    year = "2018"
}

@article{Baikov:2001aa,
    author = "Baikov, P. A. and Chetyrkin, K. G. and Kuhn, Johann H.",
    title = "{The Cross section of $e^+ e^-$ annihilation into hadrons of order $\alpha_s^4 n_f^2$ in perturbative QCD}",
    eprint = "hep-ph/0108197",
    archivePrefix = "arXiv",
    reportNumber = "FREIBURG-THEP-01-13, TTP-01-19",
    doi = "10.1103/PhysRevLett.88.012001",
    journal = "Phys. Rev. Lett.",
    volume = "88",
    pages = "012001",
    year = "2002"
}

@article{Baikov:2008jh,
    author = "Baikov, P. A. and Chetyrkin, K. G. and Kuhn, Johann H.",
    title = "{Order $\alpha_s^4$ QCD Corrections to $Z$ and $\tau$-Decays}",
    eprint = "0801.1821",
    archivePrefix = "arXiv",
    primaryClass = "hep-ph",
    reportNumber = "SFB-CPP-08-04, TTP08-01",
    doi = "10.1103/PhysRevLett.101.012002",
    journal = "Phys. Rev. Lett.",
    volume = "101",
    pages = "012002",
    year = "2008"
}

@article{Baikov:2010je,
    author = "Baikov, P. A. and Chetyrkin, K. G. and Kuhn, J. H.",
    title = "{Adler Function, Bjorken Sum Rule, and the Crewther Relation to Order $\alpha^4_s$ in a General Gauge Theory}",
    eprint = "1001.3606",
    archivePrefix = "arXiv",
    primaryClass = "hep-ph",
    reportNumber = "TTP10-05",
    doi = "10.1103/PhysRevLett.104.132004",
    journal = "Phys. Rev. Lett.",
    volume = "104",
    pages = "132004",
    year = "2010"
}

@article{Baikov:2012er,
    author = "Baikov, P. A. and Chetyrkin, K. G. and Kuhn, J. H. and Rittinger, J.",
    title = "{Complete ${\cal O}(\alpha_s^4)$ QCD Corrections to Hadronic $Z$-Decays}",
    eprint = "1201.5804",
    archivePrefix = "arXiv",
    primaryClass = "hep-ph",
    reportNumber = "SFB-CPP-12-05, TTP11-31",
    doi = "10.1103/PhysRevLett.108.222003",
    journal = "Phys. Rev. Lett.",
    volume = "108",
    pages = "222003",
    year = "2012"
}

@article{Baikov:2012zm,
    author = "Baikov, P. A. and Chetyrkin, K. G. and Kuhn, J. H. and Rittinger, J.",
    title = "{Vector Correlator in Massless QCD at Order $\mathcal{O}(\alpha^4_s)$ and the QED beta-function at Five Loop}",
    eprint = "1206.1284",
    archivePrefix = "arXiv",
    primaryClass = "hep-ph",
    reportNumber = "SFB-CPP-12-36, TTP12-018",
    doi = "10.1007/JHEP07(2012)017",
    journal = "JHEP",
    volume = "07",
    pages = "017",
    year = "2012"
}

@article{Baikov:2005rw,
    author = "Baikov, P. A. and Chetyrkin, K. G. and Kuhn, Johann H.",
    title = "{Scalar correlator at $O(\alpha_s^4)$, Higgs decay into $b$-quarks and bounds on the light quark masses}",
    eprint = "hep-ph/0511063",
    archivePrefix = "arXiv",
    reportNumber = "SFB-CPP-05-33, TTP05-11",
    doi = "10.1103/PhysRevLett.96.012003",
    journal = "Phys. Rev. Lett.",
    volume = "96",
    pages = "012003",
    year = "2006"
}

@article{Herzog:2017dtz,
    author = "Herzog, F. and Ruijl, B. and Ueda, T. and Vermaseren, J. A. M. and Vogt, A.",
    title = "{On Higgs decays to hadrons and the R-ratio at N$^{4}$LO}",
    eprint = "1707.01044",
    archivePrefix = "arXiv",
    primaryClass = "hep-ph",
    reportNumber = "NIKHEF-2017-029, LTH-1136",
    doi = "10.1007/JHEP08(2017)113",
    journal = "JHEP",
    volume = "08",
    pages = "113",
    year = "2017"
}

@article{Djouadi:1994gf,
    author = "Djouadi, A. and Gambino, P.",
    title = "{QCD corrections to Higgs boson selfenergies and fermionic decay widths}",
    eprint = "hep-ph/9406431",
    archivePrefix = "arXiv",
    reportNumber = "UDEM-GPP-TH-94-02, NYU-TH-94-06-01",
    doi = "10.1103/PhysRevD.51.218",
    journal = "Phys. Rev. D",
    volume = "51",
    pages = "218--228",
    year = "1995",
    note = "[Erratum: Phys.Rev.D 53, 4111 (1996)]"
}

@article{Chetyrkin:2000mq,
    author = "Chetyrkin, K. G. and Steinhauser, M.",
    title = "{Three loop nondiagonal current correlators in QCD and NLO corrections to single top quark production}",
    eprint = "hep-ph/0012002",
    archivePrefix = "arXiv",
    reportNumber = "DESY-00-173, FREIBURG-THEP-00-17, TTP-00-25",
    doi = "10.1016/S0370-2693(01)00179-4",
    journal = "Phys. Lett. B",
    volume = "502",
    pages = "104--114",
    year = "2001"
}

@article{Chetyrkin:2001je,
    author = "Chetyrkin, K. G. and Steinhauser, M.",
    title = "{Heavy-light current correlators at order $\alpha_s^2$ in QCD and HQET}",
    eprint = "hep-ph/0108017",
    archivePrefix = "arXiv",
    reportNumber = "DESY-01-090, FREIBURG-THEP-01-07, TTP-01-14, THEP-01-07, TTP01-14",
    doi = "10.1007/s100520100744",
    journal = "Eur. Phys. J. C",
    volume = "21",
    pages = "319--338",
    year = "2001"
}

@article{Maier:2011jd,
    author = "Maier, A. and Marquard, P.",
    title = "{Low- and High-Energy Expansion of Heavy-Quark Correlators at Next-To-Next-To-Leading Order}",
    eprint = "1110.5581",
    archivePrefix = "arXiv",
    primaryClass = "hep-ph",
    doi = "10.1016/j.nuclphysb.2012.01.021",
    journal = "Nucl. Phys. B",
    volume = "859",
    pages = "1--12",
    year = "2012"
}

@article{Hoff:2011ge,
    author = "Hoff, Jens and Steinhauser, Matthias",
    title = "{Moments of heavy-light current correlators up to three loops}",
    eprint = "1103.1481",
    archivePrefix = "arXiv",
    primaryClass = "hep-ph",
    reportNumber = "SFB-CPP-11-05, TTP11-02",
    doi = "10.1016/j.nuclphysb.2011.04.007",
    journal = "Nucl. Phys. B",
    volume = "849",
    pages = "610--627",
    year = "2011"
}

@article{Chetyrkin:2006bj,
    author = "Chetyrkin, K. G. and Faisst, M. and Kuhn, Johann H. and Maierhofer, P. and Sturm, Christian",
    title = "{Four-Loop QCD Corrections to the Rho Parameter}",
    eprint = "hep-ph/0605201",
    archivePrefix = "arXiv",
    reportNumber = "SFB-CPP-06-24, TTP06-17",
    doi = "10.1103/PhysRevLett.97.102003",
    journal = "Phys. Rev. Lett.",
    volume = "97",
    pages = "102003",
    year = "2006"
}

@article{Boughezal:2006xk,
    author = "Boughezal, R. and Czakon, M.",
    title = "{Single scale tadpoles and ${\cal O}(G_F m_t^2 \alpha_s^3)$ corrections to the $\rho$ parameter}",
    eprint = "hep-ph/0606232",
    archivePrefix = "arXiv",
    doi = "10.1016/j.nuclphysb.2006.08.007",
    journal = "Nucl. Phys. B",
    volume = "755",
    pages = "221--238",
    year = "2006"
}

@article{Chetyrkin:2010dx,
    author = "Chetyrkin, K. G. and Maier, A.",
    title = "{Massless correlators of vector, scalar and tensor currents in position space at orders $\alpha_s^3$ and $\alpha_s^4$: Explicit analytical results}",
    eprint = "1010.1145",
    archivePrefix = "arXiv",
    primaryClass = "hep-ph",
    reportNumber = "SFB-CPP-10-89",
    doi = "10.1016/j.nuclphysb.2010.11.007",
    journal = "Nucl. Phys. B",
    volume = "844",
    pages = "266--288",
    year = "2011"
}

\end{document}